%% file: text.tex
\documentclass{PoS}
\usepackage{amsmath}

\input commands.tex

\title{Testing universality and \\ automatic O($a$) improvement \\ in massless lattice QCD  with Wilson quarks}

\ShortTitle{Testing universality and automatic O($a$) improvement}

\author{\speaker{Stefan Sint}\\
        School of Mathematics, Trinity College, Dublin 2, Ireland\\
        E-mail: \email{sint@maths.tcd.ie}}

\author{Bj\"orn Leder\\
        Fachgruppe Mathematik, Bergische Universit\"at, Gau{\ss}stra{\ss}e 20, D-42097 Wuppertal, Germany\\
        E-mail: \email{leder@math.uni-wuppertal.de}}

\abstract{The chirally rotated Schr\"odinger functional provides a test bed
for universality and automatic O(a) improvement. We here report on  extensive quenched simulations of lattice QCD 
with Wilson quarks in the massless limit. We demonstrate that, after proper tuning of a dimension 3 boundary counterterm, 
the expected chirally rotated boundary conditions are indeed obtained. This implies automatic O($a$) improvement 
which we then verify in a few examples. Universality of properly renormalized correlation functions
is confirmed by comparing to the standard set-up of the Schr\"odinger functional. 
As a by-product of this study the non-singlet current renormalisation 
constants $Z_{\rm A}$ and $Z_{\rm V}$ are obtained from ratios of 2-point functions.}

\FullConference{The XXVIII International Symposium on Lattice Field Theory, Lattice2010\\
                June 14-19, 2010\\
                Villasimius, Italy}

\begin{document}

\section{Introduction}

The Schr\"odinger functional (SF) is a useful tool to solve renormalization problems in lattice gauge 
theories~[1--4]. 
A drawback of the standard formulation of the SF for Wilson fermions~\cite{Sint:1993un} 
consists in large bulk O($a$) effects even in the massless limit. While these O($a$) effects can be cancelled by
the usual Sheikholeslami-Wohlert term and O($a$) counterterms to the composite fields~\cite{Sheikholeslami:1985ij,Luscher:1996sc}, 
it is surprising that these bulk O($a$) counterterms are required at all, given the fact that 
Wilson fermions at zero mass enjoy the property of automatic O($a$) improvement~\cite{Frezzotti:2003ni,Sint:2005qz}. 
The origin of this problem is the explicit breaking of chiral symmetry
by the standard SF boundary conditions for the fermions. Rendering automatic O($a$) 
improvement compatible with SF-like boundary conditions has been 
shown to be possible for theories with even flavour
numbers~\cite{Sint:2005qz,Sint:2010eh}.
An attractive solution are chirally rotated SF boundary conditions, which, 
for $\Nf=2$ flavours take the form
\begin{equation}
  \tilde{Q}_+\psi(x)\vert_{x_0=0} =0 = \tilde{Q}_-\psi(x)\vert_{x_0=T}, \qquad
  \psibar(x)\tilde{Q}_+\vert_{x_0=0} = 0 = \psibar(x)\tilde{Q}_-\vert_{x_0=T}, 
  \label{eq:bc_rot}
\end{equation}
with $\tilde{Q}_\pm=\frac12(1\pm i\gamma_0\gamma_5\tau^3)$ and Pauli matrices $\tau^{1,2,3}$ 
acting in flavour space.  The non-trivial flavour structure 
implies that $\gamma_5\tau^1$ commutes with $\tilde{Q}_\pm$, and can be
used to resurrect the argument of automatic O($a$) improvement.
Chirally rotated SF boundary conditions derive their name from the fact that
they arise from the standard SF boundary conditions 
by performing a non-anomalous chiral rotation of the flavour doublet fields,
\begin{equation}
    \psi\rightarrow R(\alpha)\psi,\qquad \psibar \rightarrow \psibar R(\alpha), \qquad 
     R(\alpha)=\exp(i\alpha \gamma_5\tau^3/2).
 \label{eq:chiral_rot}
\end{equation}
The rotated fields satisfy boundary conditions involving the projectors  
$P_\pm(\alpha)=\frac12\left[1\pm \gamma_0\exp(i\gamma_5\tau^3)\right]$, 
which interpolate between the standard SF boundary conditions ($P_\pm(0)\equiv P_\pm$) and the chirally 
rotated ones in Eq.~(\ref{eq:bc_rot}), as $P_\pm(\pi/2)=\tilde{Q}_\pm$.
Since chiral rotations are  symmetries of the massless QCD bulk action, one may derive 
universality relations between correlation functions calculated at different
values of $\alpha$,
\begin{equation}
 \left\langle O[\psi,\psibar]\right\rangle_{(\tilde{Q}_{\pm})}=
 \left\langle O[R(-\pi/2)\psi,\psibar R(-\pi/2)]\right\rangle_{(P_\pm)}.
 \label{eq:mapping}
\end{equation}
Here we have indexed the correlation functions by the projectors appearing in the boundary conditions.
On the lattice with Wilson quarks and the standard Wilson bulk action, one expects to recover such 
universality relations between appropriately renormalised correlation functions in the continuum limit.
However, note that with Wilson quarks it is a nontrivial matter to implement the chirally rotated boundary
conditions (\ref{eq:bc_rot}),  as it requires the fine tuning of a dimension 
3 boundary counterterm (s.~below). 
In this contribution we would like to address three questions: first, how difficult
is it to implement the boundary conditions (\ref{eq:bc_rot})?  Second, given such an
implementation, can we confirm relations following from  universality, Eq.~(\ref{eq:mapping})? 
And finally, is automatic O($a$) improvement indeed realised?

\section{Lattice set-up}
The lattice set-up is taken from \cite{Sint:2010eh}, with the fermion part of the  action, 
\begin{equation}
   S_f = a^4\sum_{x_0=0}^T\sum_{\bfx} \psibar(x)({\cal D}_W +\delta {\cal D}_W +m_0)\psi(x),
\end{equation}
and the Wilson-Dirac operator:
\begin{equation}
 a{\cal D}_W\psi(x)= \begin{cases}        
         -U_0(x)P_-\psi(x+a\hat{\bf 0}) + (K+i\gamma_5\tau^3P_-)\psi(x), 
	        & \text{if $x_0=0$,} \\
            -U_0(x)P_-\psi(x+a\hat{\bf 0})   + K\psi(x)  - U_0(x-a\hat{\bf 0})^\dagger P_+ \psi(x-a\hat{\bf 0}), 
	        & \text{if $0<x_0<T$,}\\
        (K+i\gamma_5\tau^3P_+)\psi(x)- U_0(x-a\hat{\bf 0})^\dagger P_+ \psi(x-a\hat{\bf 0}),
                                       &    \text{for $x_0=T$.}\\
                      \end{cases}
\label{eq:wilson_dirac}
\end{equation}
The time diagonal operator $K$ is defined by
\begin{eqnarray}
 K\psi(x) &=& \left(1+ \frac12\sum_{k=1}^3\left\{ a(\nabla^{}_k+\nabla^\ast_k)\gamma_k
   -a^2\nabla^\ast_k\nabla^{}_k\right\}\right)\psi(x) 
 \mbox{} +\csw \frac{i}{4} a \sum_{\mu,\nu=0}^3\sigma_{\mu\nu}\hat{F}_{\mu\nu}(x)\psi(x).
\end{eqnarray}
Finally, the counterterm action is specified by
\begin{eqnarray}
 \delta {\cal D}_W \psi (x)  &=& \left(\delta_{x_0,0}+\delta_{x_0,T}\right)
\Bigl[\left(z_f-1\right)+ \left(d_s-1\right) a{\bf D}_s\Bigr]\psi(x),
\label{eq:deltaDW}
\end{eqnarray}
where the operator ${\bf D}_s$ should reduce to $\sum_{k=1}^3\gamma_k D_k$ in the continuum limit~\cite{Sint:2010eh}.
With our conventions the tree-level coefficients are
given by $z_f^{(0)}=1$ and $d_s^{(0)}=1/2$. While we set $d_s=d_s^{(0)}$ throughout, the 
finite renormalisation constant $z_f$ must be determined non-perturbatively.

\section{Definition of correlation functions}

We need correlation functions for both the standard and the chirally rotated SF. 
In the standard SF we follow the conventions used in the literature~\cite{Sint:1997jx}
by defining
\begin{equation}
  \fx(x_0)= -\frac12\left\langle X^{f_1f_2}(x){\cal O}_{5}^{f_2f_1}\right\rangle_{(P_\pm)},\qquad
  \ky(x_0)= -\frac16\sum_{k=1}^3 \left\langle Y_k^{f_1f_2}(x){\cal O}_{k}^{f_2f_1}\right\rangle_{(P_\pm)}.
\end{equation}
Here the fields $X^a$ and $Y^a_k$ stand for the quark bilinear fields,
\begin{equation}
X = A_0,V_0,S,P,\qquad Y_k=A_k,V_k,T_{k0},\tilde{T}_{k0}.
\end{equation}
which are defined as usual, e.g.~$A_\mu^{f_1f_2} = \psibar_{f_1}\gamma_\mu\gamma_5 \psi_{f_2}$.
We also use the boundary-to-boundary correlators,
\begin{equation}
   f_1= - \frac12 \left\langle {\cal O}_{5}^{f_1f_2} {\cal O}_{5}^{'f_2f_1}\right\rangle_{(P_\pm)},\qquad
   k_1= - \frac16\sum_{k=1}^3\left\langle {\cal O}_{k}^{f_1f_2} {\cal O}_{k}^{'f_2f_1}\right\rangle_{(P_\pm)}.
\end{equation}
For the chirally rotated SF we define correlation functions in the same way.
Since the boundary conditions distinguish up and down type flavours we keep track of the flavour
assignments by a superscript to the correlation functions. In order to avoid diagrams with disconnected fermion lines, 
we imagine a setup with 4 flavours, such that there are 2 up-type flavours and 2 down-type flavours.
This greatly increases the flexibility when performing a chiral rotation, which can either rotate
two flavours of the same type or two flavours of different types, while avoiding any Wick contractions
corresponding to diagrams with disconnected fermion lines.
The correlation functions in the chirally rotated set-up are denoted by $\gx$ and $\ly$, as well as $g_1$ and $l_1$.
Their definition is such that universality implies the following relations from Eq.~(\ref{eq:mapping}):
\begin{alignat}{3}
 \fa &=  \ga^{uu'} = -i\gv^{ud}, &\qquad    \fp  &= i\gs^{uu'} =   \gp^{ud} ,&\qquad 
          \kv  &=  \lv^{uu'} =   -i\la^{ud},  \label{eq:univ1}\\
 \kt &= i\ltt^{uu'} =   \lt^{ud}, &\qquad   f_1  &=  g_1^{uu'} =    g_1^{ud},  &\qquad 
           k_1 &=  l_1^{uu'} =   l_1^{ud}.     \label{eq:univ2}
\end{alignat}
Here the flavour indices correspond to the quark bilinear operator being inserted, and the conventions
for the quark boundary fields are taken from ref.~\cite{Sint:2010eh}. All remaining correlation functions, such as
$\fv$ or $\gv^{uu'}$ are expected to vanish by parity and flavour symmetries (as defined in the standard SF
basis).

\section{Numerical simulations and results}

We have carried out a quenched simulation measuring the correlation functions for 
both the standard SF and chirally rotated SF on the same gauge configurations 
with vanishing gauge boundary fields. Both unimproved and non-perturbatively O($a$) improved Wilson quarks 
were used~\cite{Luscher:1996ug}. The chosen lattice sizes were $(L/a)^4$ with $L/a=8,12,16,24,32$. 
The physical size of the lattice was kept fixed in terms of Sommer's scale $r_0$~\cite{Sommer:1993ce}, $L/r_0=1.436$, 
using the parameterisation~\cite{Necco:2001xg} and the interpolation~\cite{Guagnelli:2002ia}.
The critical mass $\mcr$ was set by requiring the PCAC mass to vanish in the standard SF.
The simulation code is a customized version of M.~L\"uscher's DDHMC code~\cite{Luscher:2005rx}, 
and the simulations were run on PC clusters and on a BlueGene/L system.
\begin{figure}
\includegraphics[width=.49\textwidth]{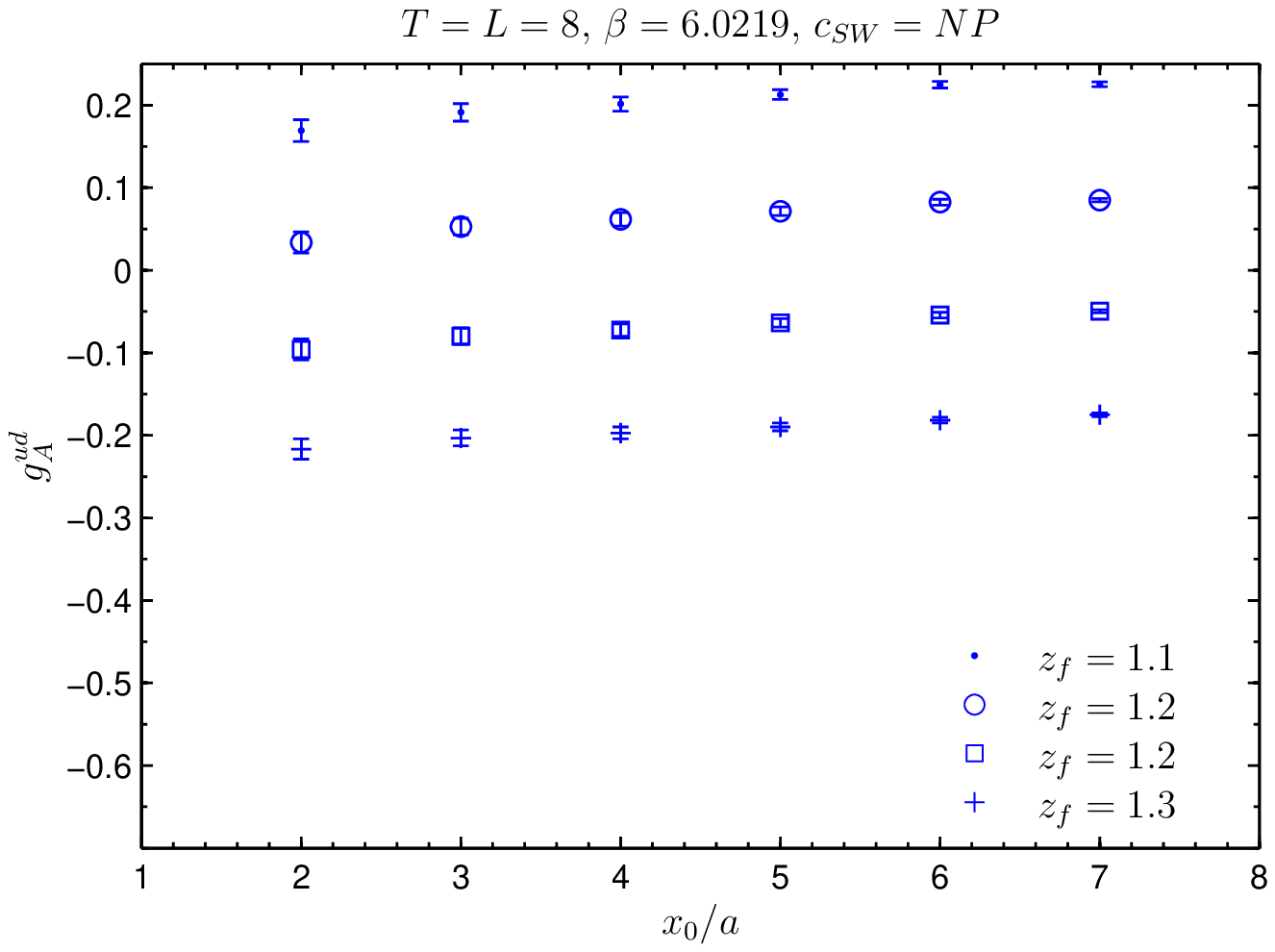}
\includegraphics[width=.49\textwidth]{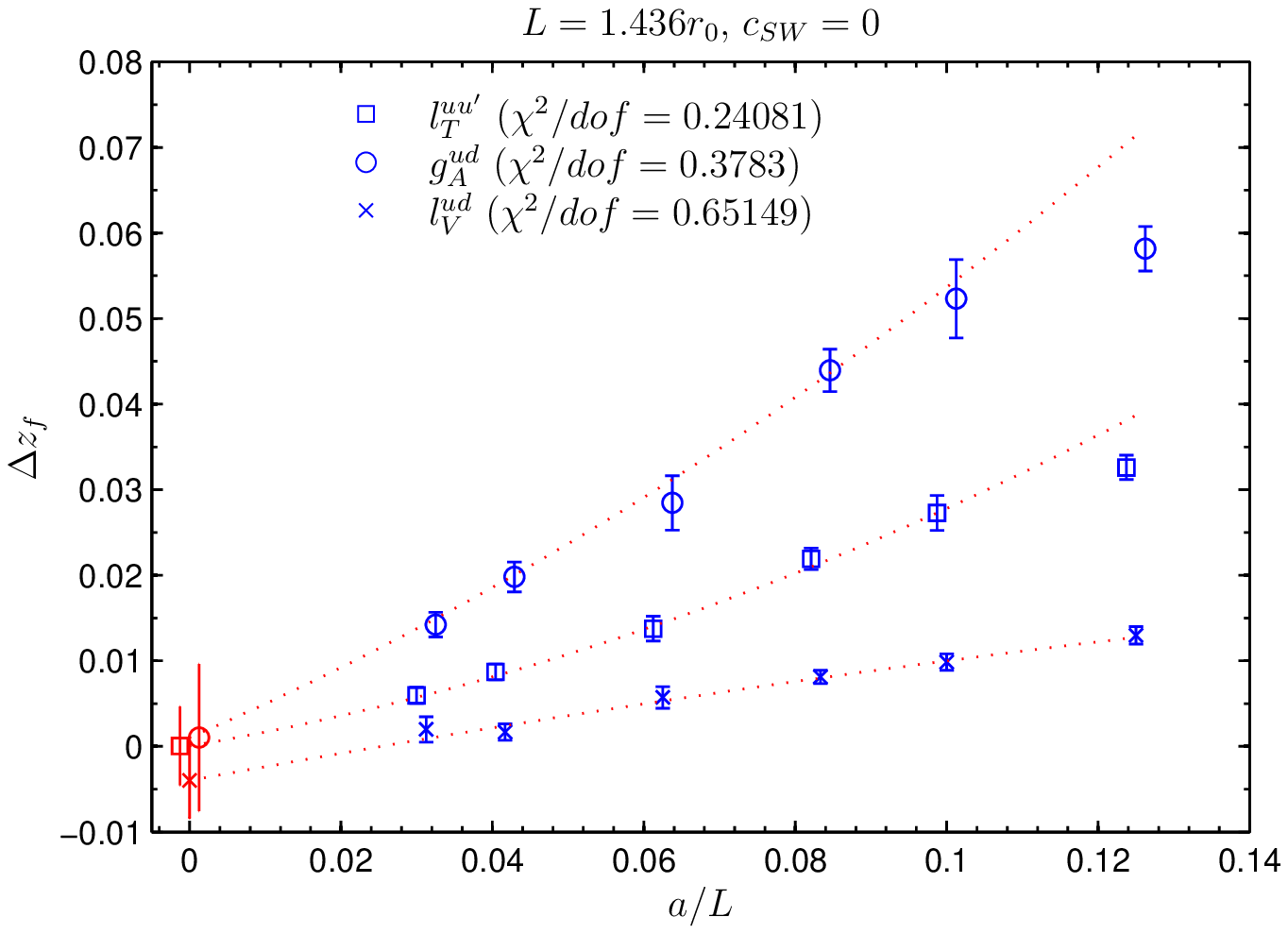}
\caption{Left panel: varying $z_f$ does not change the slope of $\ga^{ud}(x_0)$, 
which is proportional to $m_{\mathrm{PCAC}}$.  Right panel: different tuning conditions for $z_f$ yield 
O($a$) differences.}
\label{fig1}
\end{figure}

\subsection{Tuning of $z_f$, boundary conditions}

\begin{figure}
\includegraphics[width=.49\textwidth]{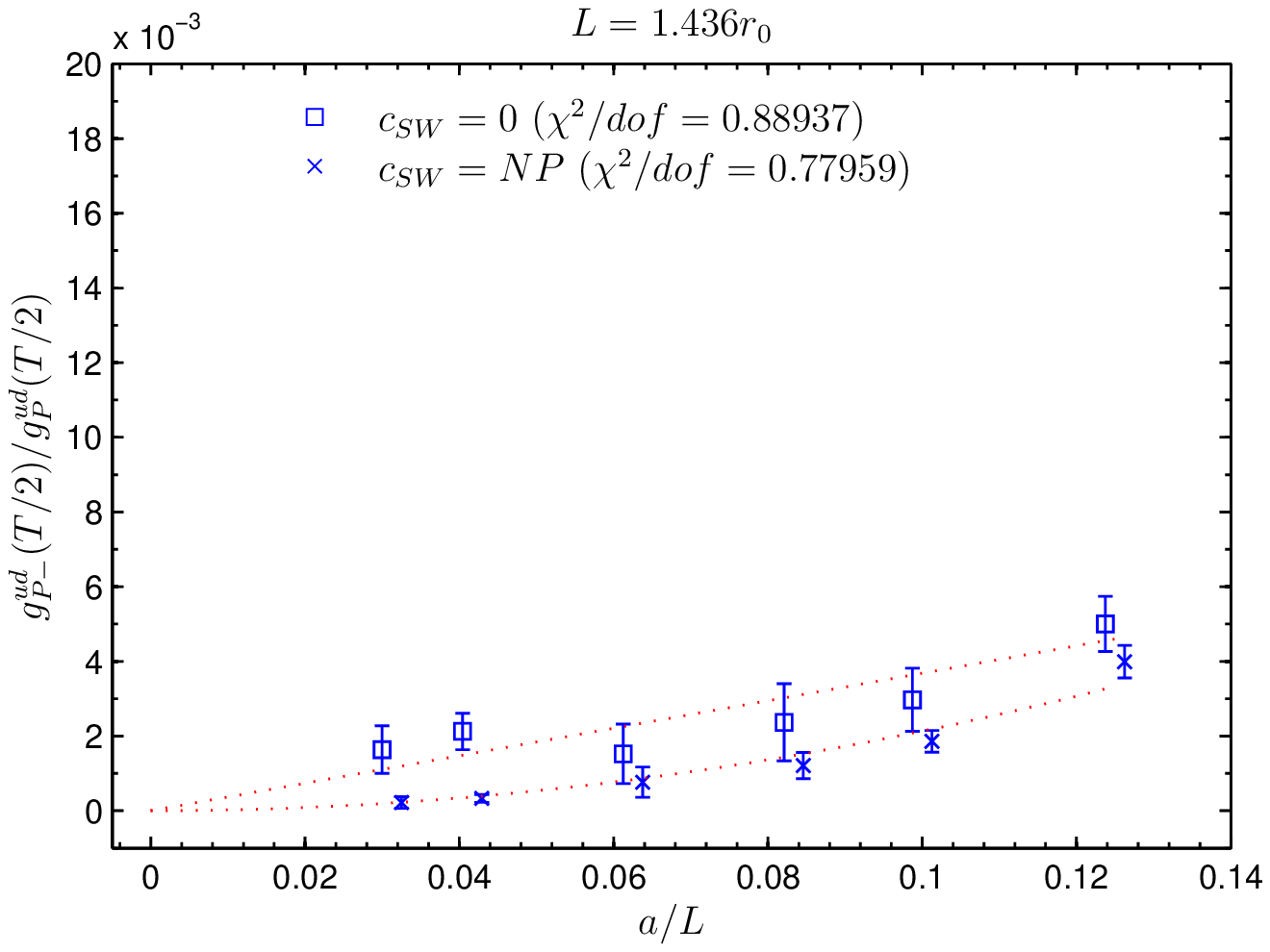}
\includegraphics[width=.49\textwidth]{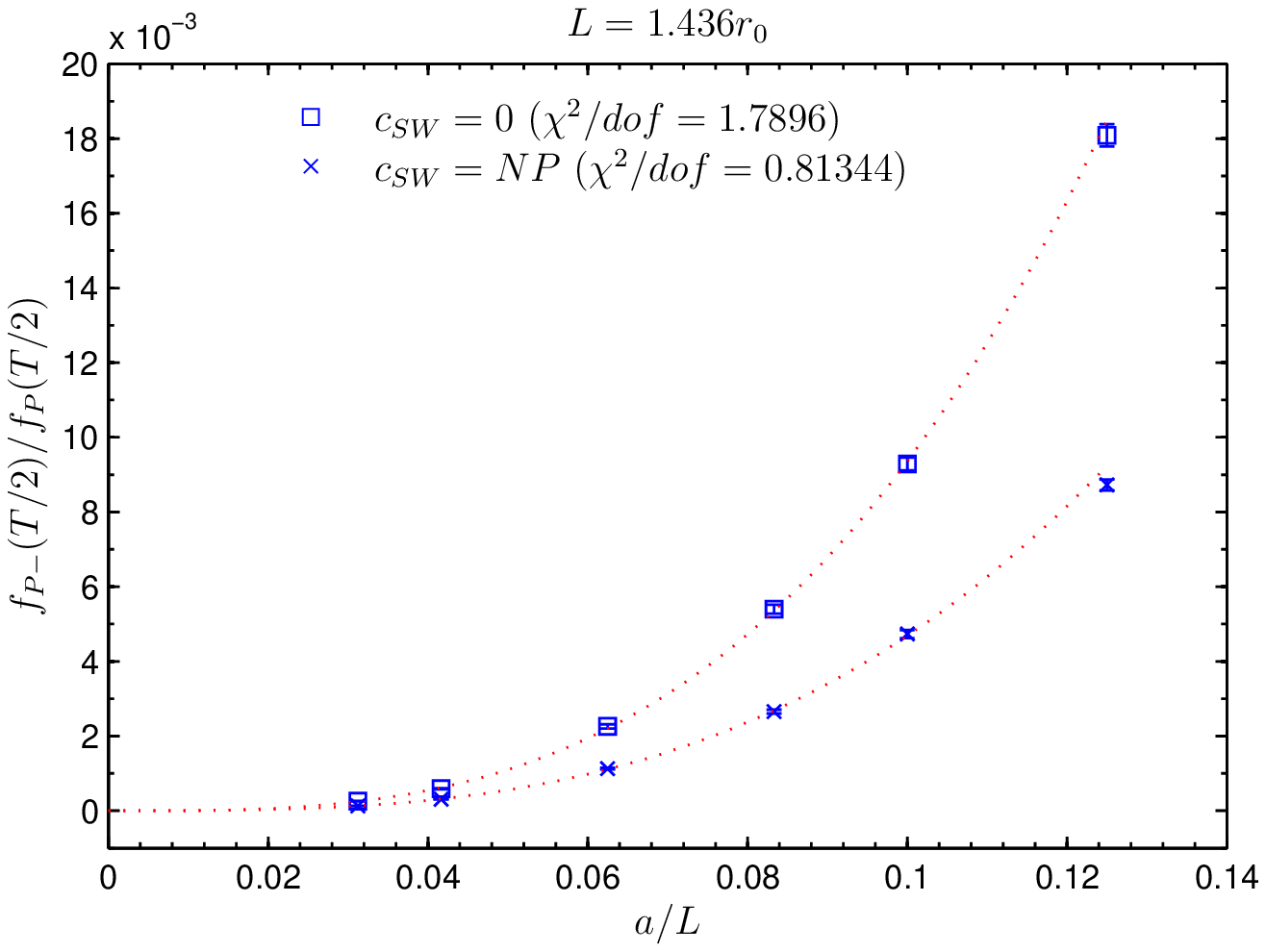}
\caption{Check of boundary conditions, cf.~main text for an explanation.}
\label{fig2}
\end{figure}

The tuning of $z_f$ can be performed by requiring any $\gamma_5\tau^1$-odd quantity
to vanish~\cite{Sint:2010eh}. Examples for such quantities are $\gp^{uu'}(x_0)$, or $\ga^{ud}(x_0)$.
Fortunately, the tuning of the parameters $m_0$ and $z_f$ is straightforward, as the respective 
tuning conditions are almost independent of each other (cf.~also~\cite{Lopez:2009yc}). 
Given the critical value $m_{\rm cr}$ of $m_0$, we have also checked that the differences $\Delta z_f$
between $z_f$ values obtained from different conditions vanish with a rate $\propto a$, as expected 
(cf.~fig.~\ref{fig1}). Given $z_f(g_0)$ one then expects that the boundary conditions 
(\ref{eq:bc_rot}) are correctly implemented up to cutoff effects. To test this hypothesis 
we have reverted the projectors $\tilde{Q}_\pm \rightarrow \tilde{Q}_\mp$ in the boundary sources and indicate
this change by a subscript "$-$" to the correlation functions. In the left panel of fig.~\ref{fig2} one indeed observes 
that the effect is very small and decreases towards zero. 
The corresponding result for the standard SF is very similar and given in the right panel of fig.~\ref{fig2}.

\subsection{Universality relations}

\begin{figure}
\includegraphics[width=.49\textwidth]{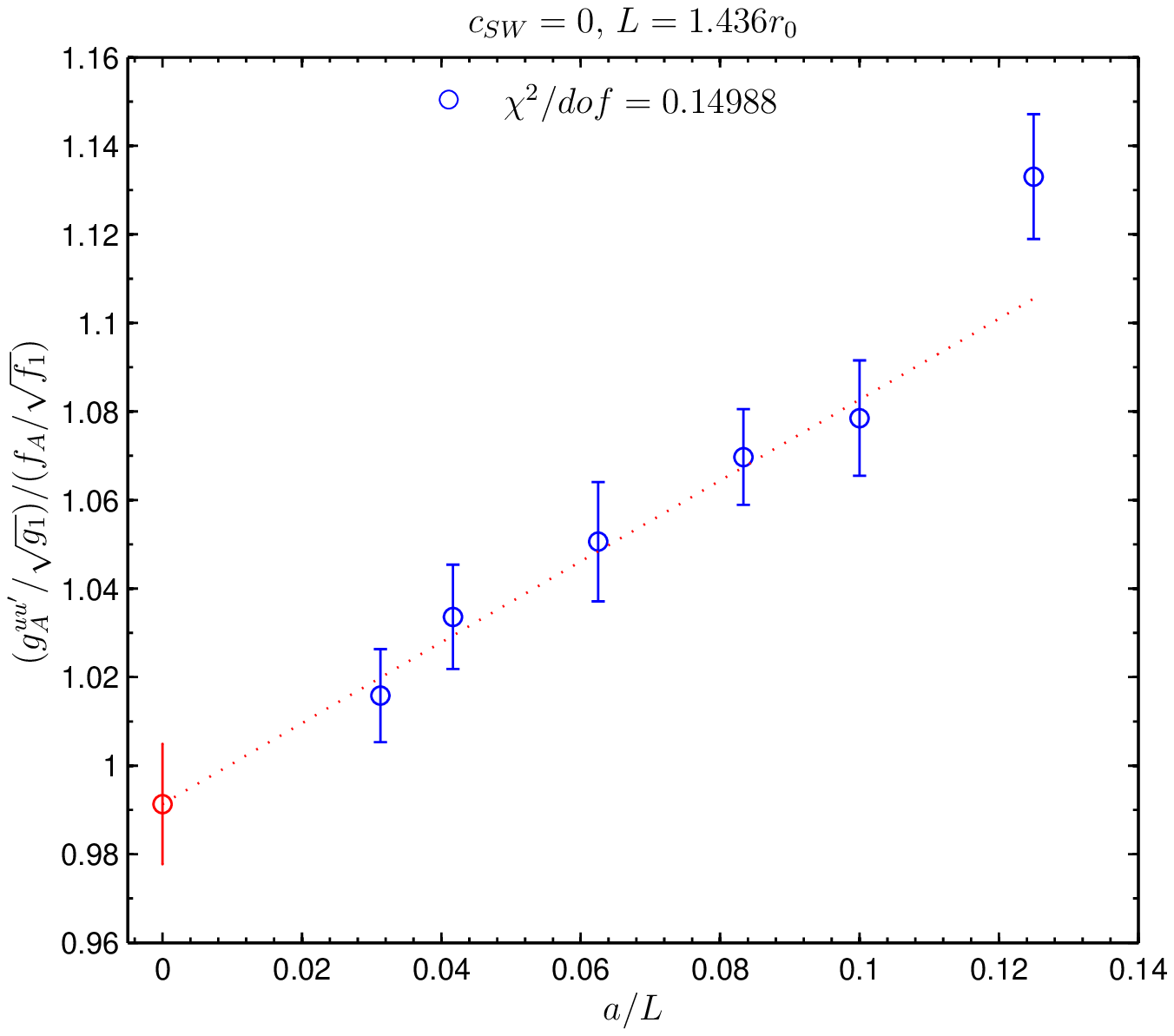}
\includegraphics[width=.49\textwidth]{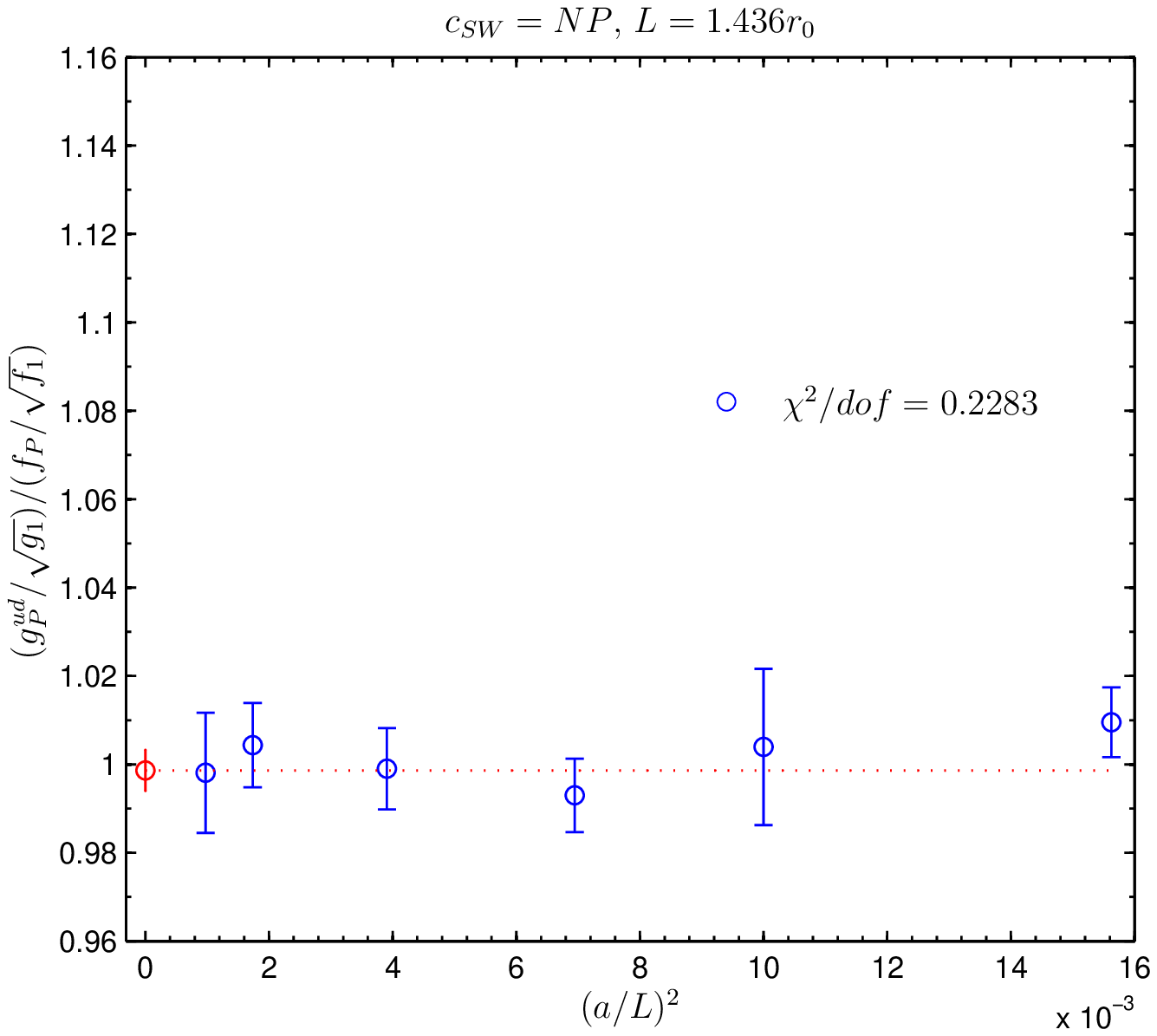}
\caption{Universality checks: we consider ratios which should approach unity in the continuum limit.}
\label{fig3}
\end{figure}

The universality relations~(\ref{eq:univ1}),(\ref{eq:univ2}) are expected to hold between renormalized correlation functions.
For instance one should then find that the ratios
\begin{equation}
  \left[\ga^{uu'}(T/2)/\sqrt{g_1})\right]\times\left[\fa(T/2)/\sqrt{f_1}\right]^{-1},\qquad
  \left[\gp^{ud}(T/2)/\sqrt{g_1})\right]\times\left[\fp(T/2)/\sqrt{f_1}\right]^{-1},
\end{equation}
approach unity in the continuum limit. As seen in fig.~\ref{fig3}, this is well satisfied within errors.

\subsection{Automatic O($a$) improvement}

\begin{figure}
\includegraphics[width=.49\textwidth]{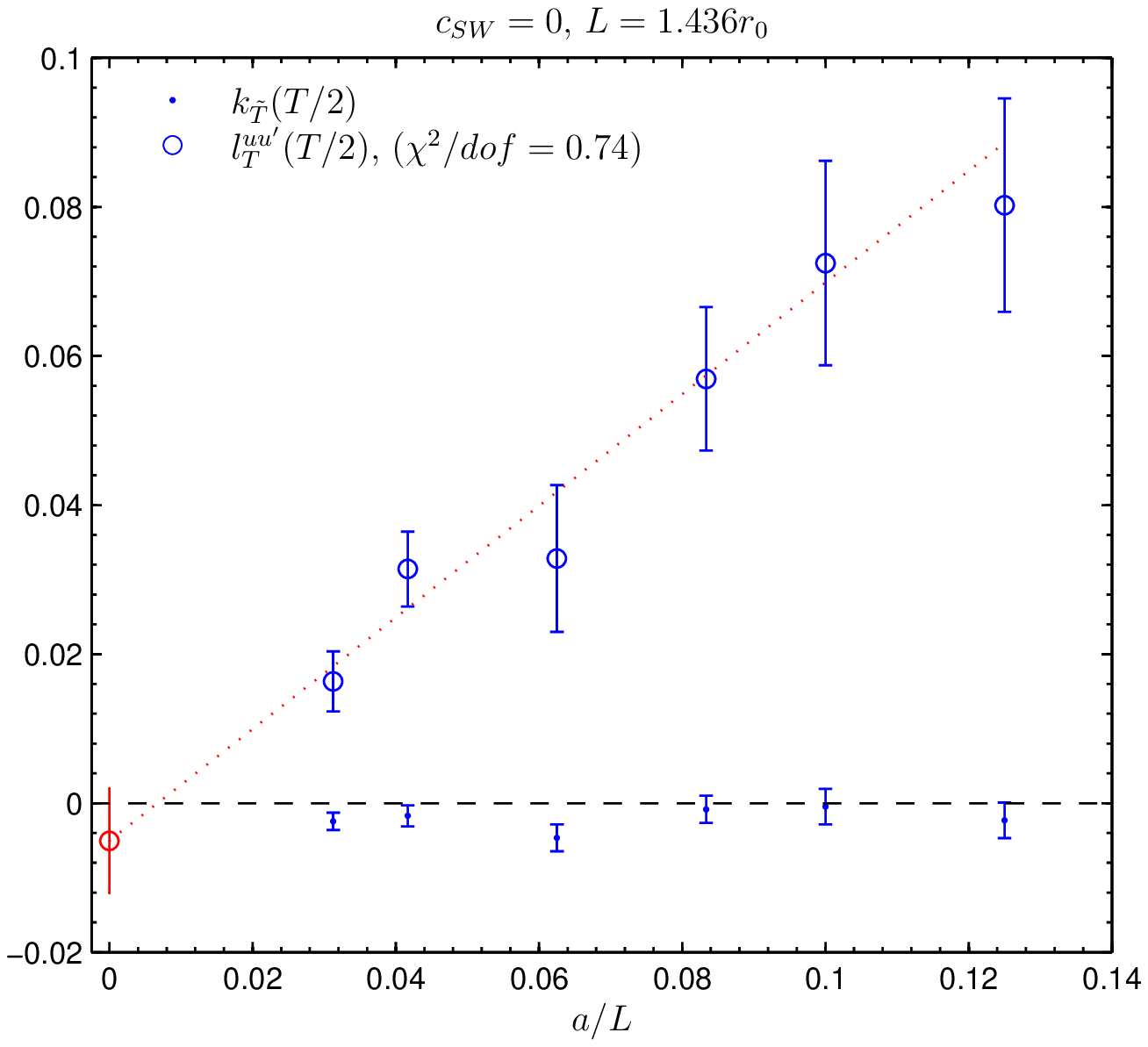}
\includegraphics[width=.49\textwidth]{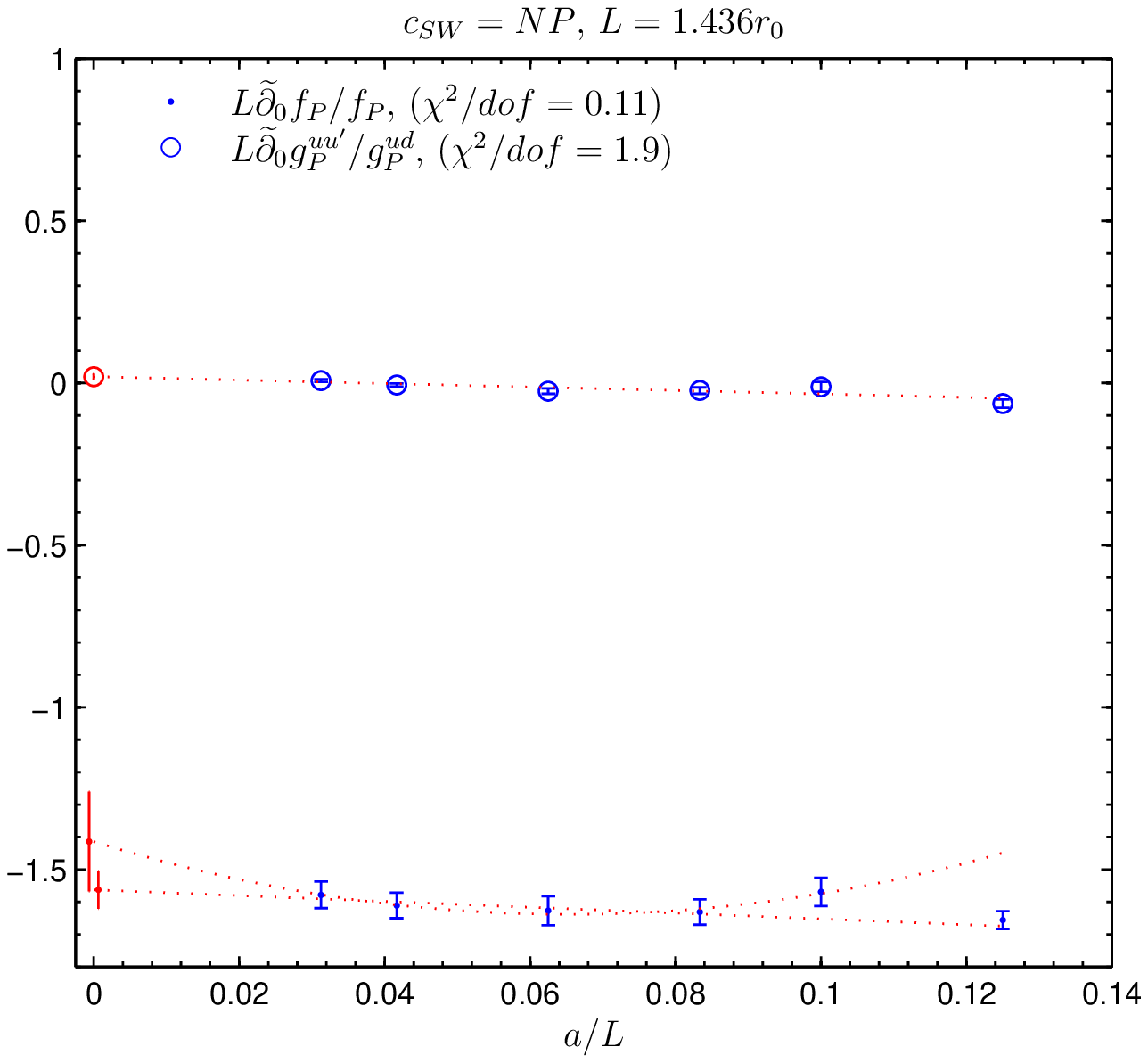}
\caption{Check of automatic O($a$) improvement, cf.~main text for details.} 
\label{fig4}
\end{figure}

Automatic bulk O($a$) improvement relies on the fact that the bulk O($a$) effects
are located in  $\gamma_5\tau^1$-odd correlation functions. By projecting on the $\gamma_5\tau^1$-even 
correlators one thus gets rid of the bulk O($a$) effects. Here we study the $\gamma_5\tau^1$-odd correlators
and verify that these vanish in the continuum limit with a rate $\propto a$, as can be seen in fig.~\ref{fig4}
for $\lt^{uu'}(T/2)$. The corresponding standard SF correlator is also shown and vanishes exactly after gauge average,
as expected due to parity and flavour symmetries. Another example is the 
counterterm contribution $\propto \ca (a/L)$  needed to improve 
correlation functions of the axial current~\cite{Luscher:1996sc}. 
The result in this case is shown in the right panel of fig.~\ref{fig4}. 
While the continuum limit vanishes in the chirally rotated SF, it is finite in the standard SF. 
Hence its contribution to axial current correlators is of O($a^2$) and O($a$), respectively, 
thereby confirming the expectation from automatic O($a$) improvement.

\subsection{Determination of $Z_{\rm A,V}$}

\begin{figure}
\includegraphics[width=.49\textwidth]{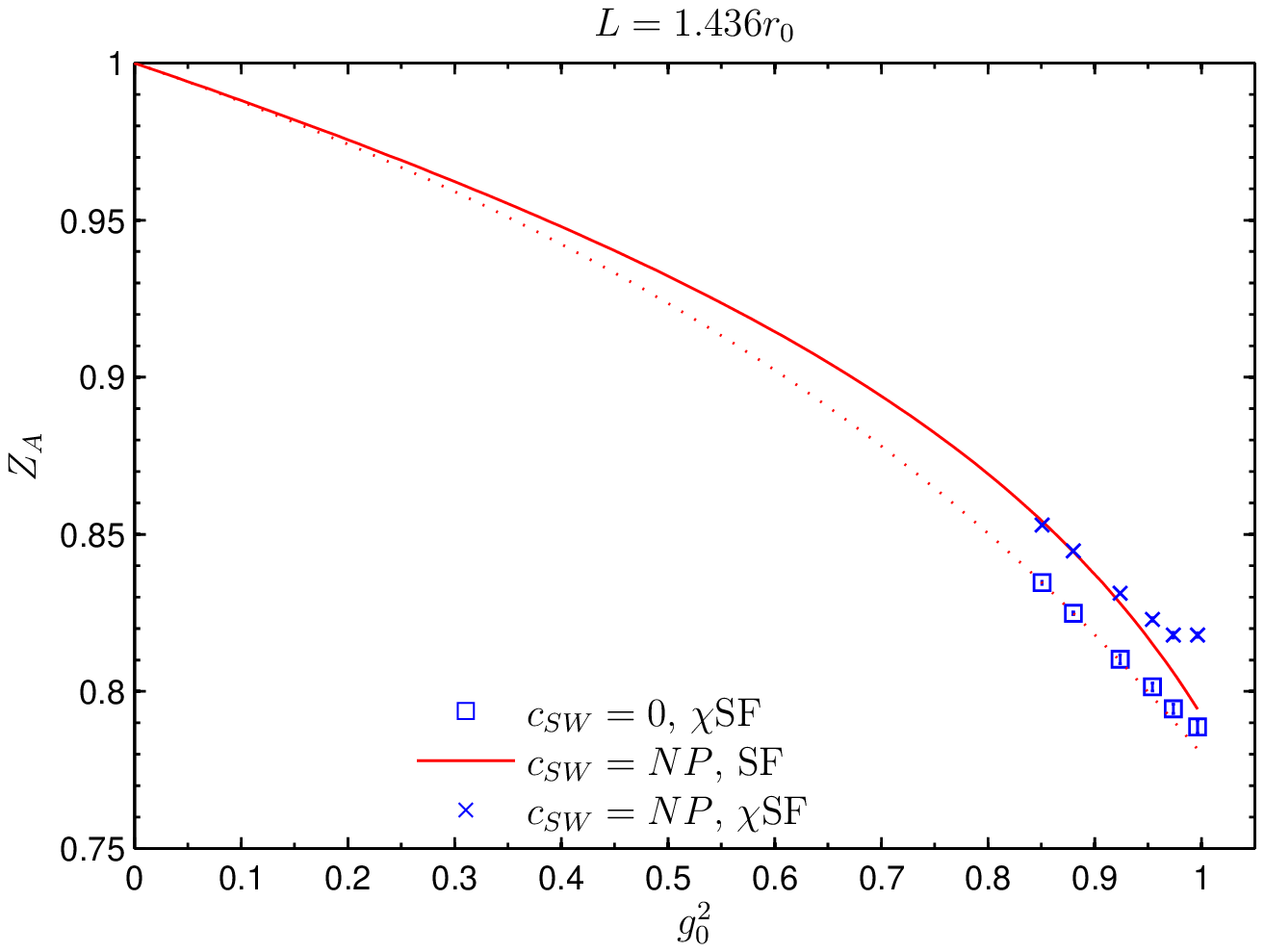}
\includegraphics[width=.49\textwidth]{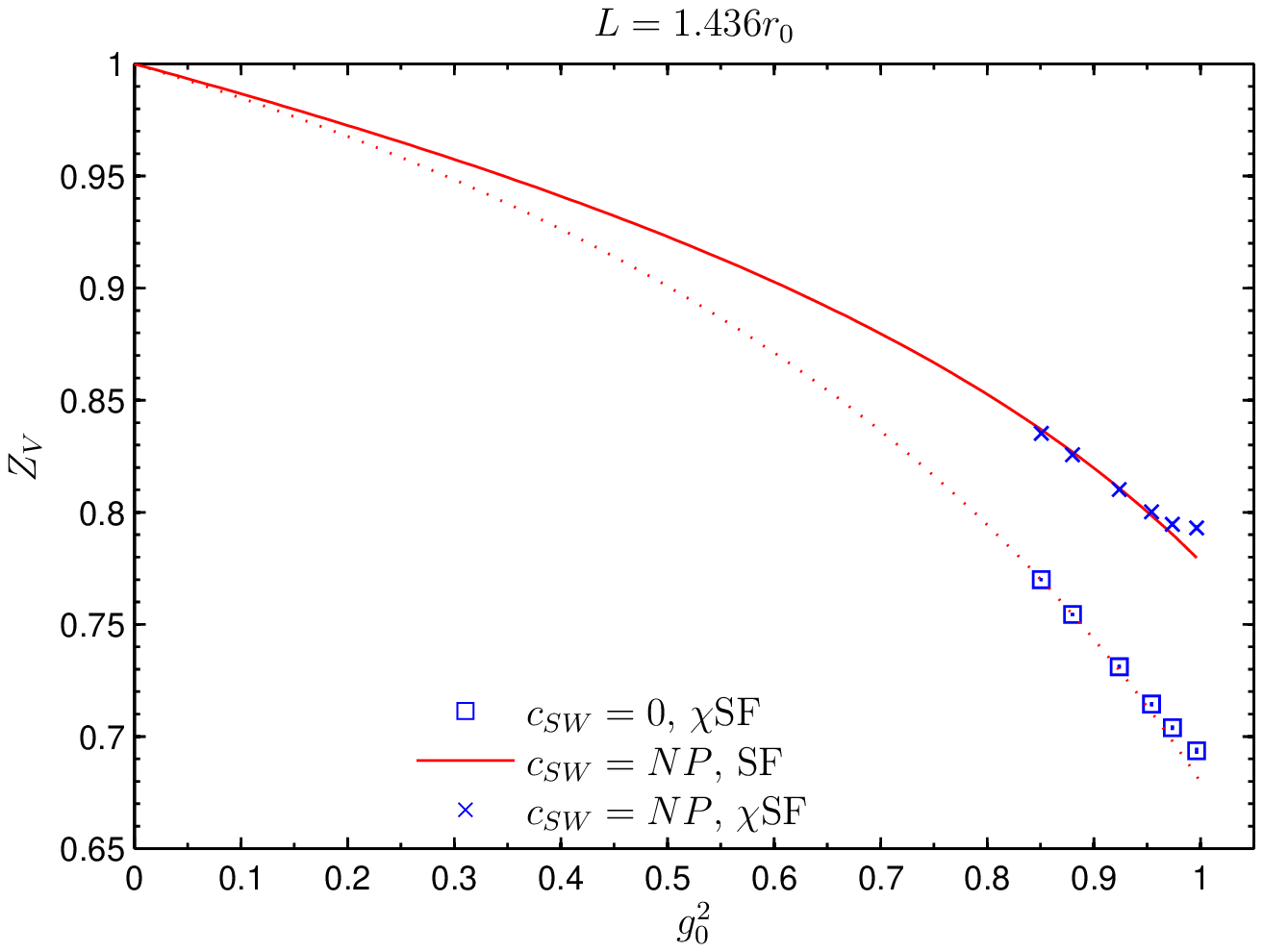}
\caption{Determination of current normalization constants. The solid line is 
the fit function from the Ward identity determination~\cite{Luscher:1996jn} with
non-perturbatively tuned $\csw$~\cite{Luscher:1996ug}.}
\label{fig5}
\end{figure}

Having checked universality we may turn the tables and use universality to determine
a number of finite renormalization constants which are usually determined from chiral Ward identities.
For instance, the continuum relations (\ref{eq:univ1}) imply that $Z_{\rm A,V}$ can be determined by
the ratios, 
\begin{equation}
 Z_{\rm A} = i g^{ud}_{\bar{\rm V}}(T/2)/\ga^{uu'}(T/2),\qquad
 Z_{\rm V} =   g^{ud}_{\bar{\rm V}}(T/2)/\gv^{ud}(T/2),
\end{equation}
where $\bar{V}_\mu$  denotes the conserved vector current. The results in fig.~\ref{fig5} are subject to
an O($a^2$) uncertainty, which perfectly explains the discrepancy with the Ward identity results
of ref.~\cite{Luscher:1996jn}.

\section{Conclusions}
We have presented a successful implementation of the chirally rotated Sch\"odinger functional.
Universality could be confirmed by comparing with standard SF correlation functions and automatic
O($a$) improvement has been verified. In the future, we expect this framework to yield better controlled
continuum extrapolations for step-scaling functions. Furthermore it provides 
new methods to determine finite renormalizaiton constants and O($a$) improvement coefficients
for gauge theories with Wilson-type fermions.

\subsection*{Acknowledgments}
The numerical simulations have been performed at the Trinity Centre for High Performance Computing 
and the Irish Centre for High End Computing. We thank both institutions for their support. B.L.~has been
partially supported by Science Foundation Ireland under grant 06/RFP/PHY061.
Funding by the EU unter Grant Agreement number PITN-GA-2009-238353 (ITN STRONGnet) is gratefully
acknowledged.

\input refs.tex
\end{document}

%% file: commands.tex
\def\lvec#1{\setbox0=\hbox{$#1$}
    \setbox1=\hbox{$\scriptstyle\leftarrow$}
    #1\kern-\wd0\smash{
    \raise\ht0\hbox{$\raise1pt\hbox{$\scriptstyle\leftarrow$}$}}
    \kern-\wd1\kern\wd0}
\def\rvec#1{\setbox0=\hbox{$#1$}
    \setbox1=\hbox{$\scriptstyle\rightarrow$}
    #1\kern-\wd0\smash{
    \raise\ht0\hbox{$\raise1pt\hbox{$\scriptstyle\rightarrow$}$}}
    \kern-\wd1\kern\wd0}
\def\lrvec#1{\setbox0=\hbox{$#1$}
    \setbox1=\hbox{$\scriptstyle\leftrightarrow$}
    #1\kern-\wd0\smash{
    \raise\ht0\hbox{$\raise1pt\hbox{$\scriptstyle\leftrightarrow$}$}}
    \kern-\wd1\kern\wd0}

\newcommand{\mcr}{m_{\rm cr}}

\newcommand{\unit}{1\kern-.25em {\rm l}}

\newcommand{\be}{\begin{equation}}
\newcommand{\ee}{\end{equation}}
\newcommand{\bd}{\begin{displaymath}}
\newcommand{\ed}{\end{displaymath}}
\newcommand{\bea}{\begin{eqnarray}}
\newcommand{\eea}{\end{eqnarray}}
\newcommand{\ba}{\begin{array}}
\newcommand{\ea}{\end{array}}

\newcommand{\Real}{\relax{\mathsf{\Gamma\kern-.35em R}}}
\newcommand{\Int}{\relax{\mathsf{Z\kern-.40em Z}}}

\newcommand{\Nf}{N_{\rm f}}

\newcommand{\bfx}{{\bf x}}

\newcommand{\psibar}{\bar{\psi}}

\newcommand{\zetabar}{\bar{\zeta}}
\newcommand{\zetaprime}{\zeta\kern1pt'}
\newcommand{\zetabarprime}{\zetabar\kern1pt'}

\newcommand{\msbar}{{\rm \overline{MS\kern-0.05em}\kern0.05em}}

\def\ca{c_{\rm A}}

\def\csw{c_{\rm sw}}

\def\fx{f_{\rm X}}

\def\ky{k_{\rm Y}}
\def\fv{f_{\rm V}}
\def\fa{f_{\rm A}}

\def\fp{f_{\rm P}}
\def\fx{f_{\rm X}}

\def\gv{g_{\rm V}}
\def\ga{g_{\rm A}}
\def\gs{g_{\rm S}}
\def\gp{g_{\rm P}}
\def\gx{g_{\rm X}}

\def\kv{k_{\rm V}}
\def\kt{k_{\rm T}}

\def\ky{k_{\rm Y}}

\def\la{l_{\rm A}}
\def\lv{l_{\rm V}}

\def\lt{l_{\rm T}}
\def\ltt{l_{\rm \tilde{T}}}
\def\ly{l_{\rm Y}}